\documentclass[reprint,showpacs,aps,unsortedaddress]{revtex4-1}
\usepackage{graphicx,amsmath,amssymb,amsfonts,comment}
\usepackage[text={7in,9.2in},centering]{geometry}
\newcommand{\mb}{\mathbf}
\newcommand{\eeq}{\end{equation}}
\newcommand{\beq}{\begin{equation}}
\newcommand{\bea}{\begin {eqnarray}}
\newcommand{\eea}{\end {eqnarray}}

\begin{document}
\title{ Reply to ``Comment on `Nonlocal quartic interactions  and universality classes in perovskite manganites'\ "}
\author{Rohit Singh} \email{rohit.singh@iitg.ernet.in}\affiliation{Department of Physics, Indian Institute of Technology Guwahati, Guwahati 781 039, India.}
\author{Kishore Dutta} \email{kdkishore77@gmail.com}\affiliation{Department of Physics, Handique Girls' College, Guwahati 781 001, India}
\author{Malay K. Nandy} \email{mknandy@iitg.ernet.in}\affiliation{Department of Physics, Indian Institute of Technology Guwahati, Guwahati 781 039, India.}
\date {31 May 2016}

\begin{abstract}
Comment [arXiv:cond-mat.stat.mech., 1602.02087v1 (2016)]  has raised questions claiming that the nonlocal model Hamiltonian presented in [Phys. Rev. E {\bf 92}, 012123 (2015)] is equivalent to the standard (short-ranged) $\Phi^4$ theory. These claims are based on a low momentum expansion of the interaction vertex that cannot be applied to  the vertex factors containing  both low and high momenta inside the loop-integrals. 
Elaborating upon the important steps of the momentum shell decimation scheme, employed in the renormalization-group calculation, we explicitly show the interplay of internal (high) and external (low) momenta determining the loop integrals for self-energy and vertex functions giving rise to corrections (to the bare parameters) different from  those of the standard (short-ranged) $\Phi^4$ theory. Employing explicit mathematical arguments, we show that this difference persists  when the range of interaction is assumed to be long (short) ranged with respect to the lattice constant (correlation-length), yielding the critical exponents as given in the original paper.   

\end{abstract}
\maketitle

In a recent paper \cite{Singh15b}, a model Hamiltonian expressed in the Fourier space as 
\[ H=\sum_{i=1}^{n}\int\frac{d^d  k}{(2\pi)^{d}}\frac{c_0\mathbf k^{2}+r_0}{2}|\phi_{i}(\mathbf k)|^{2}\]\[+\sum_{i=1}^{n}\sum_{j=1}^{n}\int\int\int\frac{d^d k_1}{(2\pi)^{d}}\frac{d^d k_2}{(2\pi)^{d}}\frac{d^d k_3}{(2\pi)^{d}}\frac{\lambda_0}{{\left[(-\mb k_1-\mb k_2)^{2}+m^{2}\right]}^{\sigma}}\]\beq\times\phi_{i}(\mathbf k_1)\phi_{i}(\mathbf k_2)\phi_{j}(\mathbf k_3)\phi_{j}(-\mathbf k_1-\mathbf k_2-\mathbf k_3), \label{eq:HFourier}\eeq
was investigated via a renormalization-group (RG)  analysis. This model Hamiltonian was constructed by modifying  the Ginzburg-Landau (GL) Hamiltonian by incorporating  a nonlocal interaction $u(\mathbf k)=\frac{\lambda_0}{{\left[\mathbf k^{2}+m^{2}\right]}^{\sigma}}$ in the  quartic term, where $\lambda_0$ is the coupling constant and  $m$ is the screening parameter.  Nonzero lattice constant $a$ imposes an ultraviolet cutoff $\Lambda(\sim a^{-1})$ to the momentum integrations . Wilson's momentum shell decimation scheme was employed at one-loop order and the critical exponents were calculated in the leading order of $\epsilon=d_c-d$, where the critical dimension turned out to be $d_c=4+2\sigma$. It was found that the critical exponents for various values of $\sigma$ for $n=3$ in three dimensions with small screening were in good agreement with experimental estimates for a wide range of experimental perovskite manganite samples.

Commenting on this work, Diehl \cite{Diehl16} has raised questions mainly  related  to the range of interaction and on the RG calculation. He argues that the model  belongs to the universality class of the standard (short-ranged) $\Phi^4$ model. To focus upon these questions, here we make the RG calculation more transparent and also discuss upon the comments raised. 

In the RG calculation of Ref.\ \cite{Singh15b}, we followed the conventional procedure of Wilson's momentum shell decimation RG scheme \cite{Ma76} where the fields are decomposed into fast and slow modes. The fast modes $\phi_i^>(\mb k)$ (belonging to the momentum shell $\Lambda/b\leqslant k\leqslant\Lambda$)  are integrated out and the effect of this elimination on the slow modes $\phi^<_i(\mb k)$ (in the  range $0< k<\Lambda/b$) is reflected via  changes in the bare parameters $r_0$, $c_0$ and $\lambda_0$. This yields the quadratic part of the Hamiltonian involving the remaining modes as
\beq \sum_{i=1}^{n}\int\frac{d^d  k}{(2\pi)^{d}}\left[\frac{c_0\mathbf k^{2}+r_0}{2}+\Sigma_a(\mb 0)+\Sigma_b(\mb k)\right]|\phi_{i}^{<}(\mathbf k)|^{2}\label{eq:selfenergy}\eeq In the expression (\ref{eq:selfenergy}), the terms involving $\Sigma_a (\mb 0)$ and $\Sigma_b (\mb k)$ correspond to the self-energy diagrams Figs. 1(a) and 1(b) of the original paper \cite{Singh15b} where
\beq \Sigma_{a}(\mathbf 0)=2n\frac{\lambda_0}{m^{2\sigma}}\int_{q=\Lambda/b}^{\Lambda}\frac{d^d q}{(2\pi)^{d}}\frac{1}{c_0q^2+r_0} \label{eq:sigmaa} \eeq and
\beq \Sigma_{b}(\mathbf k)=4\lambda_0\int_{q=\Lambda/b}^{\Lambda}\frac{d^d q}{(2\pi)^{d}}\frac{1}{[(-\mb k-\mb q)^2+m^2]^{\sigma}}\frac{1}{c_0q^2+r_0}. \label{eq:sigmab}\eeq

Importantly, the internal lines in the self-energy loop diagrams (Fig.\ 1 of Ref.\ \cite{Singh15b}) of $\Sigma_a(\mb 0)$ and $\Sigma_b(\mb k)$ belong to the high momentum shell $\Lambda/b\leqslant q\leqslant\Lambda$, whereas the external legs belong to the region $0<k<\Lambda/b$. Thus, the above integrals over $q$ in Eqs.\ (\ref{eq:sigmaa}) and (\ref{eq:sigmab}) are restricted in the high momentum shell $\Lambda/b\leqslant q\leqslant\Lambda$ and, therefore, $q\sim \Lambda$. Since large scale (small $k$) properties of the remaining modes determine the critical behavior, it is customary to expand the self-energy correction $\Sigma_b(\mb k)$ in the limit $q\gg k$. This is also equivalent to an expansion about $\mb k=0$ where $\mb k$ is the external momentum. Thus
\beq \Sigma_b(\mb k)=\Sigma_b(\mb 0)+k_i\left(\frac{\partial\Sigma_b (\mb k)}{\partial k_i}\right)_{\mb k=0}+\frac{k_ik_j}{2!}\left(\frac{\partial^2\Sigma_b(\mb k)}{\partial k_i\partial k_j}\right)_{\mb k=0}+\ldots \label{eq:k0}\eeq with 
\beq\Sigma_b(\mb 0)=4\lambda_0\int_{q=\Lambda/b}^{\Lambda}\frac{d^d q}{(2\pi)^{d}}\frac{1}{[q^2+m^2]^{\sigma}}\frac{1}{c_0q^2+r_0},\label{eq:sigmabtay}\eeq 

\beq k_i\left(\frac{\partial\Sigma_b (\mb k)}{\partial k_i}\right)_{\mb k=0}=0,\eeq because $\mb k\cdot\mb q$ appears inside the integration and 

\begin{widetext}
\beq \frac{k_ik_j}{2!}\left(\frac{\partial^2\Sigma_b(\mb k)}{\partial k_i\partial k_j}\right)_{\mb k=0}=k^2\int_{q=\Lambda/b}^{\Lambda}\frac{d^d q}{(2\pi)^d}\left[\frac{2\sigma(\sigma+1) q^2\cos^2\theta}{(q^2+m^2)^{\sigma+2}}-\frac{\sigma}{(q^2+m^2)^{\sigma+1}}\right]\frac{1}{(c_0q^2+r_0)}=k^2 \Sigma_b''(\mb 0) \mbox {(say), {}}\label{eq:sigmabprmtay}\eeq
\end{widetext}
 where $\theta$ is the angle between the momenta $\mb k$ and $\mb q$. Thus, the quadratic part of the Hamiltonian for the remaining modes can be rearranged as 
\beq \hspace{-0.9cm}\sum_{i=1}^{n}\int\frac{d^d k}{(2\pi)^{d}}\frac{[\{c_0+2\Sigma_b''(\mb 0)\}k^{2}+\{r_0+2\Sigma_a(\mb 0)+2\Sigma_b(\mb 0)\}]}{2}|\phi_{i}^{<}(\mathbf k)|^{2}
\eeq 
so that the corrections to the bare parameters $r_0$ and $c_0$ are \beq r_0+\Delta r= r_0+2\Sigma_a(\mb 0)+2\Sigma_b(\mb 0),\label{eq:delr}\eeq 
and \beq c_0+\Delta c= c_0+2\Sigma_b''(\mb 0)\label{eq:delc}\eeq respectively.

Since $m$ is a parameter in the theory, one has the freedom to choose its magnitude as $m^{-1}\gg a$, where a is the lattice constant related to the ultraviolet cutoff $\Lambda$ as $\Lambda\sim a^{-1}$. As in the self-energy integrals [Eqs.\ (\ref{eq:sigmabtay}) and (\ref{eq:sigmabprmtay})]  the internal momentum $q\sim \Lambda$, therefore $q\gg m$. Consequently, the vertex function  $\frac{1}{(q^2+m^2)^{\sigma}}$ appearing in the expressions for $\Sigma_b(\mb 0)$ and $\Sigma_b''(\mb 0)$ cannot be expanded in a low momentum expansion $q\ll m$ as suggested in Comment \cite{Diehl16}. Thus the vertex function $(q^2+m^2)^{-\sigma}$ occurring inside these integrals are expanded only in a high momentum expansion in the limit $q\gg m$, given by $(q^2+m^2)^{-\sigma}=q^{-2\sigma}[1-\frac{\sigma m^2}{q^2}+O(\frac {m^4}{q^4})]$. This yields \beq \Sigma_b(\mb 0)=4\lambda_0\int_{q=\Lambda/b}^{\Lambda}\frac{d^d q}{(2\pi)^d}q^{-2\sigma}\left[1-\frac{\sigma m^2}{q^2}+\ldots\right]\frac{1}{(c_0q^2+r_0)},\label{eq:sigmab0}\eeq \hspace{0.5cm}and \[\Sigma_b''(\mb 0)=4\lambda_0\int_{q=\Lambda/b}^{\Lambda}\frac{d^d q}{(2\pi)^d}\left[\frac{2\sigma(\sigma+1)\cos^2\theta}{q^{2\sigma+2}}-\frac{\sigma}{q^{2\sigma+2}}+\ldots\right]\]\vspace{-0.5cm}\beq\times\frac{1}{(c_0q^2+r_0)}.\label{eq:sigmabprime0}\eeq

Close to the critical point, the factor $1/(c_0q^2+r_0)$ is expanded as $\frac{1}{(c_0q^2+r_0)}=\frac{1}{c_0q^2}\left[1-\frac{r_0}{c_0q^2}+\ldots\right]$. Substituting this expansion in Eqs.\ (\ref{eq:sigmaa}), (\ref{eq:sigmab0}), and (\ref{eq:sigmabprime0}), and integrating over internal momentum $q$, we obtain $\Delta r$ and $\Delta c$ as expressed  by Eqs.\ (11) and (12) in Ref. \cite{Singh15b}. We note that the correction  $\Delta r$ [given by Eqs.\ (\ref{eq:delr}), (\ref{eq:sigmaa}) and (\ref{eq:sigmabtay})] is different from that of the standard (short-ranged) $\phi^4$ theory mainly due to the occurrence  of the vertex factor $(q^2+m^2)^{-\sigma}$ inside the integral for $\Sigma_b(\mb 0)$. Moreover, unlike the standard $\phi^4$ theory, the  correction  $\Delta c$ [given by Eqs.\ (\ref{eq:delc}) and (\ref{eq:sigmabprmtay})] is nonzero  at one-loop order. Due to these differences we cannot expect our nonlocal model  to yield the results of the standard (short-ranged) $\phi^4$ theory. The same conclusion will be reached in a field theoretic calculation where the integrals over internal momentum $q$ in $\Sigma_a(\mb 0)$, $\Sigma_b(\mb 0)$, $\Sigma_b''(\mb 0)$ would extend from $0$ to $\infty$ and an ultraviolet pole near $q\rightarrow\infty$ should be picked up.

To calculate the vertex corrections we follow the same procedure of eliminating the fast modes $\phi_i^>(\mb k)$ belonging to the high momentum shell $\Lambda/b\leqslant k\leqslant \Lambda$. This yields the quartic part of the Hamiltonian involving the remaining modes as 

\begin{widetext} 
\[\sum_{i=1}^{n}\sum_{j=1}^{n}\int_{0}^{\Lambda/b}\frac{d^d k_1d^d k_2d^d k_3}{(2\pi)^{3d}}\frac{\lambda_0}{[(-\mb k_1-\mb k_2)^2+m^2]^{\sigma}}\phi_{i}^{<}(\mathbf k_1)\phi_{i}^{<}(\mathbf k_2)\phi_{j}^{<}(\mathbf k_3)\phi_{j}^{<}(-\mathbf k_1-\mathbf k_2-\mathbf k_3)\]
\[-4n\sum_{i=1}^{n}\sum_{j=1}^{n}\\\int_{0}^{\Lambda/b}\frac{d^d k_1d^d k_2d^d k_3}{(2\pi)^{3d}}\frac{1}{[(-\mb k_1-\mb k_2)^2+m^2]^{2\sigma}}I_a(\mb k_1, \mb k_2) \phi_{i}^{<}(\mathbf k_1)\phi_{i}^{<}(\mathbf k_2)\phi_{j}^{<}(\mathbf k_3)\phi_{j}^{<}(-\mathbf k_1-\mathbf k_2-\mathbf k_3)\]
\[ -16\sum_{i=1}^{n}\sum_{j=1}^{n}\int_{0}^{\Lambda/b}\frac{d^d k_1d^d k_2d^d k_3}{(2\pi)^{3d}}  \frac{1}{[(-\mb k_1-\mb k_2)^2+m^2]^{\sigma}}I_b(\mb k_1,\mb k_2,\mb k_3)\phi_{i}^{<}(\mathbf k_1)\phi_{i}^{<}(\mathbf k_2)\phi_{j}^{<}(\mathbf k_3)\phi_{j}^{<}(-\mathbf k_1-\mathbf k_2-\mathbf k_3)\]
\beq-16\sum_{i=1}^{n}\sum_{j=1}^{n}\int_{0}^{\Lambda/b}\frac{d^d k_1d^d k_2d^d k_3}{(2\pi)^{3d}} I_c(\mb k_1, \mb k_2, \mb k_3 )\phi_{i}^{<}(\mathbf k_1)\phi_{i}^{<}(\mathbf k_2)\phi_{j}^{<}(\mathbf k_3)\phi_{j}^{<}(-\mathbf k_1-\mathbf k_2-\mathbf k_3),\label{eq:vertex}\eeq
\end{widetext}

where 
\begin{widetext}
\[I_a (\mb k_1, \mb k_2)=\lambda_0^2\int_{q=\Lambda/b}^\Lambda\frac{d^d q}{(2\pi)^d}\,\frac{1}{c_0q^2+r_0}\frac{1}{c_0(-\mb k_1-\mb k_2-\mb q)^{2}+r_0},\]
\[I_b(\mb k_1, \mb k_2, \mb k_3)=\lambda_0^2\int_{q=\Lambda/b}^\Lambda\frac{d^d q}{(2\pi)^d}\,\frac{1}{[(\mb q-\mb k_3)^2+m^2]^{\sigma}}\frac{1}{c_0q^2+r_0}\frac{1}{c_0(-\mb k_1-\mb k_2-\mb q)^{2}+r_0},\] and
\[I_c(\mb k_1, \mb k_2, \mb k_3)=\lambda_0^2\int_{q=\Lambda/b}^\Lambda\frac{d^d q}{(2\pi)^d}\frac{1}{[(-\mb k_1-\mb q)^2+m^2]^{\sigma}}\frac{1}{[(\mb q-\mb k_2)^2+m^2]^{\sigma}}\frac{1}{c_0q^2+r_0} \frac{1}{c_0(-\mb k_1-\mb k_3-\mb q)^{2}+r_0}.\]
\end{widetext}
In the expression (\ref{eq:vertex}), the terms involving $I_a(\mb k_1, \mb k_2)$, $I_b(\mb k_1, \mb k_2, \mb k_3)$, and $I_c(\mb k_1, \mb k_2,\mb k_3)$ correspond to the vertex diagrams 2(a), 2(b) and 2(c) of the original paper \cite{Singh15b}.
They are related to Eqs.\ (13), (14) and (15) of Ref.\ \cite{Singh15b} as
\beq \Pi_a(\mb k_1, \mb k_2)=-4n\left[(-\mb k_1-\mb k_2)^2+m^2\right]^{-2\sigma}I_a(\mb k_1, \mb k_2),\eeq
\beq \Pi_b(\mb k_1, \mb k_2,\mb k_3)=-16\left[(-\mb k_1-\mb k_2)^2+m^{2}\right]^{-\sigma}I_b(\mb k_1, \mb k_2, \mb k_3),\eeq
\beq \Pi_c(\mb k_1, \mb k_2, \mb k_3)=-16I_c(\mb k_1, \mb k_2, \mb k_3).\eeq
 
The first three terms of (\ref{eq:vertex}) have the common factor of $\frac{1}{[(-\mb k_1-\mb k_2)^2+m^2]^{\sigma}}$ and therefore they can be combined to yield 
\begin{widetext}
\[\sum_{i=1}^{n}\sum_{j=1}^{n}\int_{0}^{\Lambda/b}\frac{d^d k_1d^d k_2d^d k_3}{(2\pi)^{3d}}\left[\lambda_0-4n\frac{I_a(\mb k_1, \mb k_2)}{\left[(-\mb k_1-\mb k_2)^2+m^2\right]^{\sigma}}- 16I_b(\mb k_1, \mb k_2, \mb k_3)\right]\frac{1}{\left[(-\mb k_1-\mb k_2)^2+m^2\right]^{\sigma}}\]
\beq \phi_{i}^{<}(\mathbf k_1)\phi_{i}^{<}(\mathbf k_2)\phi_{j}^{<}(\mathbf k_3)\phi_{j}^{<}(-\mathbf k_1-\mathbf k_2-\mathbf k_3),\label{eq:vercomb}\eeq 
\end{widetext}
whereas the last term of (\ref{eq:vertex}) involving $I_c(\mb k_1, \mb k_2, \mb k_3)$, having no factor of the original interaction $\frac{1}{[(-\mb k_1-\mb k_2)^2+m^2]^{\sigma}}$, cannot be incorporated inside the square bracket of the above expression (\ref{eq:vercomb}). Thus the last term of Eq.\ (\ref{eq:vertex})  cannot generate a term similar to the first term of (\ref{eq:vertex}) (involving the factor $\frac{1}{[(-\mb k_1-\mb k_2)^2+m^2]^{\sigma}}$). Being unable to generate a correction to $\lambda_0$, the last term of (\ref{eq:vertex}) is thus irrelevant. Consequently, the box diagram in Fig. 2(c) does not contribute to a  correction to $\lambda_0$. 

It may be noted that the box diagram [Fig. 2(c)] is found to be irrelevant even in the absence of screening, as shown in Ref.\ \cite{Singh15a}. The vertex correction obtained in Ref.\ \cite{Singh15a} can be compared with that of Weinrib and Halperin \cite{Weinrib83} where the effective Hamiltonian involves a nonlocal quartic interaction term with coupling function $g(k)=-v-wk^{\sigma-d}$ and a term $u|\phi^{\alpha}(\mb x)|^{4}$.  Thus, their model reduces to that of Ref.\ \cite{Singh15a} for $u=v=0$ and  $w=-w$. For this choice of parameters, the  RG flow equation for the long-range vertex $w$ given in Ref.\ \cite{Weinrib83} reduces to that of Ref.\ \cite{Singh15a}.

Since in Wilson's renormalization procedure the effect of elimination of small scales (internal momentum $q$) on the large scales (external momenta $k_i$) is to be found out, the above corrections are expanded in the limit $q\gg k_i$ which is equivalent to an expansion about $\mb k_i=0$ (This is equivalent  to the renormalization-point ($\mb k_i=0$) in the field theoretic language). Thus, upon expanding $I_a(\mb k_1, \mb k_2)/\left[(-\mb k_1-\mb k_2)^2+m^2\right]^{\sigma}$ and $I_b(\mb k_1, \mb k_2, \mb k_3)$ about $\mb k_i=0$, the bare coupling constant $\lambda_0$ acquires relevant corrections as 
\beq \lambda_0+\Delta\lambda=\lambda_0-\frac{4n}{m^{2\sigma}}I_a(\mb 0, \mb 0)-16I_b(\mb 0, \mb 0, \mb 0),\label{eq:dellambda}\eeq where \beq I_a(\mb 0, \mb 0)=\lambda_0^2\int_{q=\Lambda/b}^\Lambda\frac{d^d q}{(2\pi)^d}\,\frac{1}{(c_0q^2+r_0)^2},\label{eq:ia0}\eeq \beq I_b (\mb 0, \mb 0, \mb 0)=\lambda_0^2\int_{q=\Lambda/b}^\Lambda\frac{d^d q}{(2\pi)^d}\,\frac{1}{[q^2+m^2]^{\sigma}}\frac{1}{(c_0q^2+r_0)^2}. \label{eq:ib}\eeq

As discussed before, since $m$ is a parameter in the theory, one has the freedom to choose a long-range interaction (that is a large value for the range $m^{-1}$) compared to the lattice constant $a$. Consequently, we choose  $m^{-1}\gg a$. Since the ultraviolet cutoff $\Lambda\sim a^{-1}$, therefore $m\ll\Lambda$. As the integrals over $q$ in the expressions for $I_a(\mb 0, \mb 0)$ and $I_b(\mb 0, \mb 0, \mb 0)$ given by Eqs.\ (\ref{eq:ia0}) and (\ref{eq:ib}) are restricted in the high momentum shell $\Lambda/b\leqslant q\leqslant\Lambda$, this implies $q\sim\Lambda$ and therefore $q\gg m$. Thus the vertex factor $(q^2+m^2)^{-\sigma}$ occurring inside the integral for $I_b(\mb 0, \mb 0, \mb 0)$ can only be expanded  in a high momentum expansion in the limit $q\gg m$, given by $(q^2+m^2)^{-\sigma}=q^{-2\sigma}[1-\frac{\sigma m^2}{q^2}+O(\frac {m^4}{q^4})]$  giving \beq I_b (\mb 0, \mb 0, \mb 0)=\lambda_0^2\int_{q=\Lambda/b}^\Lambda\frac{d^d q}{(2\pi)^d}q^{-2\sigma}[1-\frac{\sigma m^2}{q^2}+\ldots]\frac{1}{(c_0q^2+r_0)^2}\label{eq:ib0}.\eeq

In contrast, Comment \cite{Diehl16} suggested a low momentum expansion in the limit $q\ll m$,  given by $\frac{1}{(q^2+m^2)^{\sigma}}=\frac{1}{m^{2\sigma}}+O(q^2)$ which cannot be employed because $\Lambda\sim q\gg m$. The same conclusion will be reached in a field theoretic RG scheme with integrations over the internal momentum $q$ ranging from $0$ to $\infty$ where the ultraviolet pole near $q\rightarrow\infty$ should be picked up \cite{ZinnJustin}.  

Near the critical point, the factor $1/(c_0q^2+r_0)$ is expanded as $\frac{1}{(c_0q^2+r_0)}=\frac{1}{c_0q^2}\left[1-\frac{r_0}{c_0q^2}+\ldots\right]$. Using this expansion in Eqs.\ (\ref{eq:ia0}) and (\ref{eq:ib0}), and performing the integrations over the internal momentum $q$ in the high momentum shell $\Lambda/b\leqslant q\leqslant\Lambda$,  we obtain the correction $\Delta\lambda$  as given by Eq.\ (16) in Ref. \cite{Singh15b}.

This vertex correction $\Delta\lambda$ together with $\Delta r$ and $\Delta c$, being different from that of the short-ranged $\phi^4$ theory, do not yield the critical exponents of the standard $\Phi^4$ theory. 
This is because of the occurrence of the nonlocal vertex function in the loop integrals for $\Sigma_b(\mb k)$ and $I_b(\mb k_1, \mb k_2, \mb k_3)$ together with the fact that $I_c(\mb k_1, \mb k_2, \mb k_3)$ does not contribute to $\Delta\lambda$. Thus, the nonlocal model Hamiltonian [Eq.\ (\ref{eq:HFourier})] does not generate the results of the standard $\Phi^4$ model such as mean-field exponents for $d>4$ and the standard critical exponents of the short-ranged $\Phi^4$ theory for $2<d<4$ dimensions (unlike what is suggested in Comment \cite{Diehl16}).  
The obtained nontrivial fixed point from the RG calculation does not yield mean-field exponents for any choice of $\sigma$ (either positive or negative), including the range $4<d<4+2\sigma$ for positive $\sigma$. Instead, only tricritical  mean-field exponents are obtained for $\sigma=-0.5$ which is equivalent to $\epsilon=d_c-d=4+2\sigma-d=0$ in three dimensions as given in Ref.\ \cite{Singh15b}.

As stated earlier, $m$ being a parameter in the theory, we have the freedom to choose its value so that the interaction  is long-ranged with respect to the lattice constant, that is $m^{-1}\gg a$ or $m\ll\Lambda$. Thus the range of interaction extends over many lattice points unlike the standard short-ranged $\phi^4$ theory with nearest-neighbour interaction.  As we have already explained, the resulting self-energy and vertex corrections turn out to be different from that of the standard $\phi^4$ theory. This situation does not change even when we approach the critical point allowing us to expand $(c_0q^2+r_0)^{-1}$ inside the integrals in the limit $\Lambda^2\gg r_0$. Since the renormalized value $r\sim\xi^{-2}$, this implies $\Lambda\gg\xi^{-1}$ or $a\ll\xi$, as expected.  Because of these limits, the integrals for the  self-energy [$\Sigma_a(\mb 0)$, $\Sigma_b(\mb 0)$, and $\Sigma_b''(\mb 0)$] and vertex [$I_a(\mb 0, \mb 0)$  and $I_b(\mb 0, \mb 0, \mb 0)$] contributions  are effectively expanded in powers of the dimensionless ratios $m/\Lambda$ and $a/\xi$. Out of these dimensionless ratios $a/\xi$ is infinitesimal near the critical point whereas $m/\Lambda$ remains finite because $m$ and $\Lambda$ are determined by the intrinsic properties of the substance such as its chemical composition and physical structure. It may be noted that the dimensionless ratio $m^{-1}/\xi$ does not occur in the expansions for the above integrals. Hence the corrections $\Delta r$, $\Delta c$, and $\Delta\lambda$ to the bare parameters $r_0$, $c_0$ and $\lambda_0$ depend  on the ratio $m/\Lambda$  as given in Ref.\ \cite{Singh15b}. These corrections being different from those of the standard $\phi^4$ theory, they yield results different from those of the standard (short-ranged) $\phi^4$ theory. As $m/\Lambda$ remains finite at the critical point and $\sigma$ is another parameter in the theory, the critical indices are found to depend on the values of $\sigma$ and $m/\Lambda$. It is found that the critical indices vary slowly with respect to the parameter $m/\Lambda$ whereas the variations with respect to $\sigma$ are significant.

In this context it is worthwhile to mention about the  known fact that critical exponents can vary with the strength of coupling  for models  with energy-energy coupling. In the Ashkin-Teller Potts model, there are two sets of Ising spins  given by the Hamiltonian $-J\sum_{nn}SS-K\sum_{nn}TT-K_4\sum_{nn}SSTT$ that yields critical exponents which vary continuously with the coupling constant $K_4$ \cite{Kadanoff79}. 
The line of critical points, instead of separated critical points, does not need a revision of our basic ideas of renormalization-group, and universality. There will still be restricted universality for a particular point  on the line of critical fixed points as all critical exponents are determined by a single parameter \cite{Kadanoff79,Dhar}.

Since the model Hamiltonian [Eq.\ (\ref{eq:HFourier})] contains the  parameters $m$ and $\sigma$, the resulting critical indices depend on them via the dimensionless quantities $\sigma$ and $m/\Lambda$, as explicitly shown by the RG calculation. As in the Ashkin-Teller Potts model, this does not require one to redefine the notion of universality. We may regard the cases with the same value of $m/\Lambda$ to belong to a particular class wherein the variation of  critical exponents with $\sigma$ to different subclasses of universality.

In our RG analysis, we followed the standard procedure of obtaining the upper critical dimension  $d_c$ from the  marginal stability of the stable eigenvalue for the nontrivial fixed point, yielding $d_c=4+2\sigma$. Consequently, an $\epsilon$-expansion with $\epsilon=d_c-d$ was carried out. It can be checked that the fixed point values $r^*$ and $\lambda^*$ turn out to be $O(\epsilon)$ in the leading order.  This is consistent with the notion of $\epsilon$-expansion in critical phenomena \cite{Ma76}. The six critical exponents are found to acquire corrections  at $O(\epsilon)$ that yield critical exponent values different from those of the standard $\Phi^4$ model.    

It is important to note that the method of power  counting also yields the same upper critical dimension. The Hamiltonian $H$ expressed in the momentum space [Eq.\ (\ref{eq:HFourier})] yields the momentum dimension of the field $[\phi_i(\mb k)]=[\Lambda^{-1-d/2}]$. As a result, the dimension of $\lambda_0$ turns out to be $[\lambda_0]=[\Lambda^{d-4-2\sigma}]$ for any value of $m$ (including $m=0$, because the dimension of $\lambda_0$ is independent of the value of $m$) giving $d_c=4+2\sigma$.

In Comment \cite{Diehl16} an example of a modified Model C Hamiltonian by inserting the operator $(m^2-\nabla^2)^\sigma$  in the $\psi^2$ term with an interaction term $\gamma_0\psi\phi^2$ is suggested. By dropping out the $\Phi^4$ term and integrating out the $\psi$ field (after treating $\nabla^2$ as irrelevant), it is stated that it belongs to the same universality class as that of the standard $+\Phi^4$ model. However, a $-\Phi^4$ term is generated upon eliminating the $\psi$ field by means of integration. As the Hamiltonian with a $-\Phi^4$
potential is unstable, it requires a $+\Phi^6$ term to stabilize it and hence it cannot be in the same universality class as that of the $+\Phi^4$ model. Moreover, the assumption that the interaction vertex $\gamma_0$ could be imaginary is unphysical because the Hamiltonian, being classical, must be a real valued quantity. This is also applicable for $m=0$ case.

The origin of nonlocality and its connection with  perovskite manganites (R$_{1-x}$A$_x$MnO$_3$) may be traced as follows. Spin-lattice coupling,  known to be important in perovskite manganites, is directly related to the imposed strain \cite{Zhao96,Babushkina98,Millis98,Millis98b,Ahn04} due to perturbations induced via  a change in  $R$, $A$, and $x$. A number of theoretical investigations \cite{Fisher68,Wagner70a,Wagner70c,Aharony73} on a compressible magnetic lattice  have shown the emergence of nonlocal quartic term in the effective Hamiltonian as a result of spin-lattice coupling. It has  been shown \cite{Aharony73}  via an RG calculation that  tricriticality emerges from such an effective Hamiltonian. Interestingly, a wide number of perovskite manganite samples exhibit tricritical behavior in addition to behavior away from tricriticality. Thus it is interesting to consider a model Hamiltonian with nonlocal interaction in the quartic term and see whether such a model can exhibit varying critical exponents including tricriticality observed in perovskite manganites.  

In an attempt to capture such a behavior, a  model with algebraic decay of interaction with infinite range ($m=0$) was considered in Ref.\ \cite{Singh15a}. An RG calculation \cite{Singh15a} with this model could indeed generate tricritical exponents  of some perovskite manganite samples satisfactorily. However, this model could not represent satisfactorily the critical exponents of those samples that are away from tricriticality. In particular, the values for the exponent $\beta$ are found to be restricted in the range $0.250\leqslant\beta\leqslant 0.375$, whereas there exist perovskite manganite samples that have $\beta$ values higher than $0.375$. Consequently, it was realised that the range of the interaction may not be infinite and a model with a finite range of interaction ($m\neq 0$) , namely, Eq.\ (\ref{eq:HFourier}), was constructed.
Quite interestingly, this latter model ($m\neq 0$) [Eq.\ (\ref{eq:HFourier})] could represent satisfactorily \cite{Singh15b} the static critical exponents of a wide range of perovskite manganite samples both near and away from tricriticality, including those exhibiting $\beta$ values higher than $0.375$.  These experimental agreements signify that the model expressed by the effective Hamiltonian [Eq.\ (\ref{eq:HFourier})] contains the universality classes of perovskite manganites. It is customary to determine the predictive power of a model by its range of predictability. In this sense, comparison with experimental data becomes important.

The authors would like to thank Professor Deepak Dhar, Tata Institute of Fundamental Research, Mumbai, for fruitful discussions.

\end{document}